\documentclass[aps,prd,twocolumn,amsmath,amssymb,superscriptaddress]{revtex4-1}
\usepackage{amsmath}
\usepackage{graphicx}
\usepackage{subfigure}
\usepackage{epstopdf}
\usepackage{color}
\usepackage{multirow}
\usepackage{setspace}
\usepackage{overpic}
\usepackage{amssymb}
\usepackage{braket}
\usepackage{makecell}
\usepackage[bookmarksnumbered, pdfstartview=FitH,colorlinks,urlcolor=blue, citecolor=blue,linkcolor=blue] {hyperref}
\usepackage{lineno}
\usepackage{bm}
\usepackage{rotating}
\usepackage[utf8]{inputenc}
\usepackage{orcidlink}
\hypersetup{bookmarksnumbered=true,pdfstartview=FitH,colorlinks=true,
  urlcolor=blue,citecolor=blue,linkcolor=blue}

\newcommand{\Br}{\mathcal B}

\newcommand{\Ree}{\operatorname{Re}}

\newcommand{\lamfun}{\lambda}
\usepackage{braket}

\begin{document}
\title{\boldmath Helicity amplitudes and angular distributions in $D\to b_1(1235)(\to\omega\pi)\ell^+\nu_\ell$}

\author{Yijia Zeng\,\orcidlink{0009-0005-3279-0304}}
\email{zengyj@ihep.ac.cn}
\affiliation{Institute of High Energy Physics,
	Chinese Academy of Sciences, Beijing 100049, China}
\affiliation{University of Chinese Academy of Sciences, Beijing 100049, China}

\author{Hailong Ma\,\orcidlink{0000-0001-9771-2802}}
\email{mahl@ihep.ac.cn}
\affiliation{Institute of High Energy Physics,
	Chinese Academy of Sciences, Beijing 100049, China}

\begin{abstract}
	
Using the helicity amplitude technique, we derive the decay intensity for the charmed meson semileptonic decays
$D\to b_1(1235)\ell^+\nu_\ell, b_1(1235)\to\omega\pi$ with $\ell=e,\mu$. Based on the theoretically predicted hadronic transition form factors, the branching fractions for $D^0\to b_1(1235)^-\ell^+\nu_\ell$ and their lepton flavor universality ratios are calculated. The predicted branching fractions of the positronic channel span $(2.6$--$8.8)\times10^{-5}$, with most theoretical predictions being consistent with the previous BESIII measurement $(7.2\pm1.8^{+0.6}_{-0.8})\times10^{-5}$ by assuming $\Br(b_1(1235)\to\omega\pi)=1$. We further explore the partial decay widths and angular distributions of $D\to b_1(1235)\ell^+\nu_\ell, b_1(1235)\to\omega\pi$ and find distinct behavior among various hadronic transition form factors. Further amplitude analysis with larger statistics is thus desired to test these theoretical predictions.

\end{abstract}
\maketitle

\section{Introduction}
\label{sec:intro}

The isovector axial-vector meson $b_1(1235)$ is conventionally assigned as a $q\bar q$ state belonging to the $^1P_1$ SU(3) octet. Since its first observation in the 1960s, it has been investigated in fixed target reactions, and more recently in decays of heavy mesons and charmonium systems~\cite{ParticleDataGroup:2026mpi}. While these studies have provided measurements of its mass and width, the internal structure of the $b_1(1235)$ remains disputed. Besides the conventional $q\bar q$ interpretation, admixture with $q\bar qg$ hybrid configuration~\cite{Ishida:1991nz} and molecular state generated by $K\bar{K}^*$ interactions~\cite{Lutz:2003fm,Roca:2005nm,Zhou:2014ila,Liang:2019vhf,Clymton:2023txd} have been proposed. Semileptonic charmed meson decays $D\to b_1(1235)\ell^+\nu_\ell$ provide an ideal platform to investigate the internal structure of $b_1(1235)$. Compared to the hadronic processes which suffer from complex interference patterns and final state interactions, the semileptonic decays are theoretically clearer since leptons are not involved in strong interaction. Numerous theoretical studies of the hadronic transition form factors of $D\to b_1(1235)$ have been preformed in the frameworks of ISGW models~\cite{Isgur:1988gb,Scora:1995ty}, covariant light-front quark models~(CLFQM)~\cite{Cheng:2003sm,Verma:2011yw,Yang:2025gfz}, and light-cone sum rules~(LCSR)~\cite{Momeni:2019uag,Huang:2021owr} under the conventional $q\bar q$ configuration. Experimental determination of these form factors is therefore highly desired to test these theoretical predictions, and pin down the nature of $b_1(1235)$.

Experimentally, using $2.93$ fb$^{-1}$ of $e^+e^-$ collision data sample taken at the center-of-mass energy $\sqrt{s}=3.773$~GeV, the BESIII collaboration has reported the first search for $D\to b_1(1235)e^+\nu_e$~\cite{BESIII:2020jan} and found no significant signal. With an enlarged data sample of $7.93$ fb$^{-1}$, BESIII subsequently reported the first observation of $D^0\to b_1(1235)^-e^+\nu_e$ and an evidence of $D^+\to b_1(1235)^0e^+\nu_e$~\cite{BESIII:2024pwp}, with a total signal yield more than $50$ events. The full $20.3$ fb$^{-1}$~\cite{BESIII:2024lbn} of data sample at $\sqrt{s}=3.773$~GeV collected by BESIII, together with the forthcoming large data sample at super $\tau$-charm factory~\cite{Cheng:2022tog}, could yield considerable statistics for both the positronic and muonic channels, thereby enabling a detailed amplitude analysis to extract the form factors.

In this paper, using the helicity amplitude technique, we derive the decay intensity for the charmed meson semileptonic decays $D\to b_1(1235)\ell^+\nu_\ell, b_1(1235)\to\omega\pi$ with $\ell=e,\mu$. By combining the derived intensity with various theoretical predictions of form factors, corresponding branching fractions, lepton flavor universality ratios, partial decay widths, and angular distributions are obtained.

\section{Helicity amplitudes}
\label{sec:amplitude}

\begin{figure}[htbp]
	\centering
	\centering
	\setlength{\abovecaptionskip}{-0.0cm}
	\includegraphics[width=0.89\linewidth]{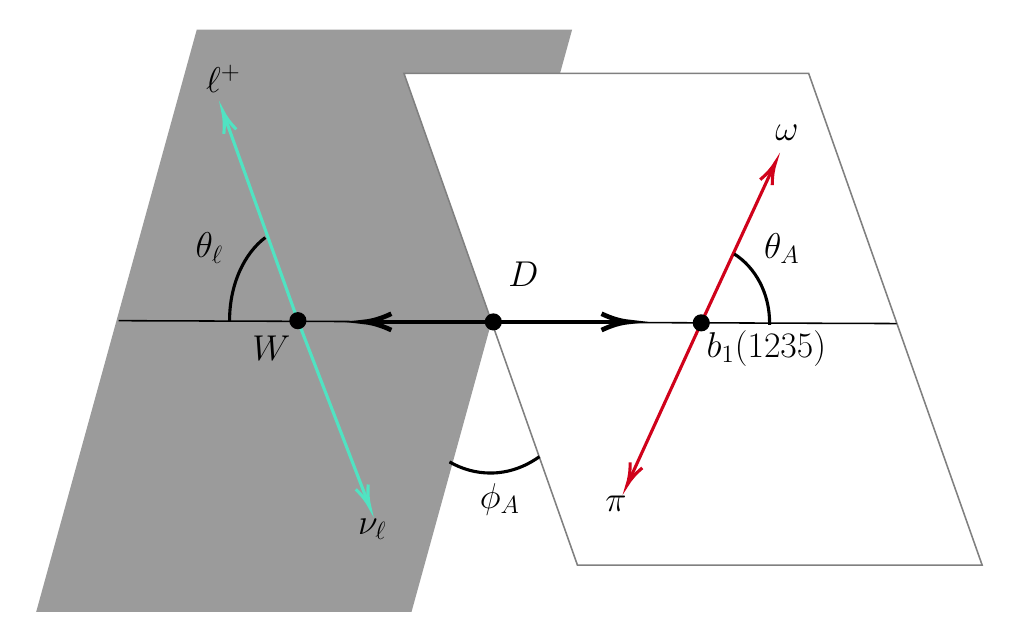}
	\caption{Definitions of the helicity angles for the $D\to b_1(1235)\ell^+\nu_\ell,b_1(1235)\to\omega\pi$ decays.}
	\label{fig:helicity_diagram}
\end{figure}

In this section, we will derive the decay intensity of $D\to b_1(1235)\ell^+\nu_\ell, b_1(1235)\to\omega\pi$ using the helicity amplitude technique. To be clear, the definitions of the helicity angles are demonstrated in Fig.~\ref{fig:helicity_diagram} and described as below:
\begin{itemize}
	\item The angle $\theta_\ell$ is defined by the $\ell^+$ flight direction and the opposite of the $D$ meson flight direction in the $\ell^+\nu_\ell$ rest frame.
	\item The angle $\theta_A$ is defined by the $\omega$ flight direction and the opposite of the $D$ meson flight direction in the $b_1(1235)$ rest frame.
	\item The angle $\phi_A$ is defined by the normal of the $\ell^+\nu_\ell$ decay plane $\hat{n}_{\ell^+\nu_\ell}=\hat{p}_{\ell^+}\times\hat{p}_{\nu_\ell}$, and the normal of the $b_1(1235)$ decay plane $\hat{n}_{b_1(1235)}=\hat{p}_{\omega}\times\hat{p}_{\pi}$ in the $D$ meson rest frame.
\end{itemize}
The squared invariant masses of $\ell^+\nu_\ell$ and $\omega\pi$ systems are defined as $q^2=M^2_{\ell^+\nu_\ell}$ and $S_A=M^2_{\omega\pi}$, respectively.

Overall, the covariant tensor amplitude formalism of $D\to b_1(1235)\ell^+\nu_\ell, b_1(1235)\to\omega\pi$ can be written as
\begin{equation}
	\mathcal{M} = \frac{G_F}{\sqrt{2}} V_{cd} H^\mu L_\mu \epsilon^{\nu}_{A} J_{\nu}\times {\rm BW}_{A}(s_A).
\end{equation}
Here, $H^\mu=\braket{A|\bar{d}\gamma^\mu(1-\gamma_5)c|D}$ is the hadronic weak current for the $D\to b_1(1235)$ transition; $L_\mu=\bar{u}_\nu\gamma_\mu(1-\gamma_5)v_{\ell}$ is the leptonic weak current for $W^*\to \ell^+\nu_\ell$; $\epsilon^{\nu}_{A} J_{\nu}$ models the subsequent strong decay $b_1(1235)\to\omega\pi$, where $\epsilon^{\nu}_{A}$ is the polarization vector of $b_1(1235)$ and $J_\nu$ is the hadronic current for the $b_1(1235)\to\omega\pi$. All three components will be described using the helicity formalism in the following subsections. The non-negligible width of $b_1(1235)$ is described with its Breit-Wigner lineshape ${\rm BW}_{A}(s_A) = {1}/{(m^2_A-s_A-im_A\Gamma_A)}$, where the resonance parameters $m_A$ and $\Gamma_A$ are fixed to their PDG averages~\cite{ParticleDataGroup:2026mpi}.

\subsection{$D\to b_1(1235) W^*$}

Following the convention in Ref.~\cite{Bian:2021gwf}, the $D\to b_1(1235)$ hadronic weak current is written as
\begin{equation}
\begin{split}
&\braket{A|\bar{d}\gamma^\mu\gamma_5 c|D} = -\frac{2iA(q^2)}{m_D-m_{A}}\epsilon^{\mu\nu\rho\sigma}(\epsilon_{A}^{*})_{\nu}(p_D)_{\rho}(p_{A})_{\sigma} \\
&\braket{A|\bar{d}\gamma^\mu c|D} = -2m_A V_0(q^2) \frac{\epsilon_A^{*}\cdot q}{q^2}q^\mu \\ &-(m_D-m_A)V_1(q^2)\left[\epsilon_A^{*\mu}-\frac{\epsilon_A^{*}\cdot q}{q^2}\,q^\mu\right] \\
&+V_2(q^2)\frac{\epsilon_A^{*}\cdot q}{m_D-m_A}\left[(p_D+p_A)^\mu-\frac{m_D^2-m_A^2}{q^2}\,q^\mu\right],
\end{split}
\end{equation}
where $q^\mu=p^\mu_{D}-p^\mu_{A}$ is the momentum transfer, $\epsilon^{\mu\nu\rho\sigma}$ is the anti-symmetric tensor, and $A(q^2)$ and $V_{0,1,2}(q^2)$ are the form factors. Hence, the helicity projections $c_{\lambda_{W}}=H^{\mu}\epsilon^*_\mu(\lambda_{W})$ are derived~\cite{Bian:2021gwf} as
\begin{equation}
\label{eq:hadronic_helicity}
\begin{split}
c_\pm &=(m_D-m_A)V_1\mp\frac{A\sqrt{\lambda(m_D^2,m_A^2,q^2)}}{m_D-m_A}, \\
c_0&=-\frac{\left[(m_D^2-m_A^2-q^2)(m_D-m_A)V_1-\frac{\lambda(m_D^2,m_A^2,q^2)}{m_D-m_A}V_2\right]}{2m_A\sqrt{q^2}} , \\
c_t&=-\frac{\sqrt{\lambda(m_D^2,m_A^2,q^2)}}{\sqrt{q^2}}V_0,
\end{split}
\end{equation}
where $\lambda(m_D^2,m_A^2,q^2)=(m^2_D+m^2_A-q^2)^2-4m_D^2m_A^2$, and $\epsilon_{W}$ is the polarization vector of $W^*$.

\subsection{$W^*\to \ell^+\nu_\ell$}

For $W^*\to \ell^+\nu_\ell$, its helicity projections $L_{\lambda_{\ell},\lambda_{W}}=L^\mu\epsilon_{\mu}(\lambda_{W})$ are given~\cite{Bian:2021gwf} as
\begin{equation}
\label{eq:weak_helicity}
\begin{array}{@{}l@{\;}l@{\quad}l@{\;}l@{}}
L_{-\frac12,-1} & = f_\ell\hat m_\ell\sin\theta_\ell,
& L_{-\frac12,0} & = -\sqrt{2}f_\ell\hat m_\ell\cos\theta_\ell,\\[3pt]
L_{-\frac12,1} & = -f_\ell\hat m_\ell\sin\theta_\ell,
& L_{-\frac12,t} & = \sqrt{2}f_\ell\hat m_\ell,\\[3pt]
L_{\frac12,-1} & = -f_\ell(1+\cos\theta_\ell),
& L_{\frac12,0} & = -\sqrt{2}f_\ell\sin\theta_\ell,\\[3pt]
L_{\frac12,1} & = -f_\ell(1-\cos\theta_\ell),
\end{array}
\end{equation}
where $f_\ell=i\sqrt{2(q^2-m_\ell^2)}$ and $\hat{m}_\ell=m_\ell/\sqrt{q^2}$.

\subsection{$b_1(1235)\to\omega\pi$}

The amplitude for the strong decay $b_1(1235)\to\omega\pi$ is constructed via the covariant tensor amplitude formalism~\cite{Zou:2002ar} as
\begin{equation}
\mathcal M_{A\to P\omega}=\epsilon^{\nu}_{A} J_{\nu}=\epsilon_A^\mu\epsilon_\omega^{*\nu}\left[g_s g_{\mu\nu}+g_d\widetilde t^{(2)}_{\mu\nu}\right].
\end{equation}
Here, $g_s$ and $g_d$ are the complex coupling constants for the $S$- and $D$-wave decays, respectively;  $\epsilon_\omega$ is the polarization vector of $\omega$; $\widetilde t^{(2)}_{\mu\nu}$ is the pure-orbital-angular-momentum covariant tensor of the $D$-wave defined in Ref.~\cite{Zou:2002ar}. A Blatt-Weisskopf factor $B_2(Q,Q_0)$ for the $D$-wave has been included in $\widetilde t^{(2)}_{\mu\nu}$, where $Q$ is the breaking momentum of $b_1(1235)\to\omega\pi$, and the hadron ``scale" $Q_0$ is chosen to be 3~GeV$^{-1}$~\cite{ParticleDataGroup:2026mpi}.

In the rest frame of $b_1(1235)$ with the $\omega$ flight direction along the $+z$ axis, the helicity projections $h_{\lambda_w}=\epsilon^\mu_A(\lambda_w)J_\mu$ are derived as
\begin{equation}
\label{eq:strongdecay_helicity}
\begin{split}
	&h_{+1}=h_{-1}\equiv h_T =-g_s-\frac43 g_d Q^2B_2(Q), \\
	&h_0\equiv h_L=\frac{E_\omega}{m_\omega}\left[-g_s+\frac83 g_d Q^2B_2(Q)\right],
\end{split}
\end{equation}
where $E_{\omega}$ is the energy of $\omega$ in the $b_1(1235)$ rest frame and $m_{\omega}$ is the nominal mass of $\omega$~\cite{ParticleDataGroup:2026mpi}.

The $S$- and $D$-wave intensity ratio $D/S$ of the $b_1(1235)\to\omega\pi$ decay has been determined in Ref.~\cite{E852:2002wsj} to be $|D/S|=0.269\pm0.014$ and $\arg(D/S)=(10.5\pm4.6)^{\circ}$. Further conversion from the helicity basis to the $LS$ basis gives the relationship
\begin{equation}
\begin{split}
	h_T &= \sqrt{\frac{1}{3}}S + \sqrt{\frac{1}{6}} D, \\
	h_L &= \sqrt{\frac{1}{3}}S - \sqrt{\frac{2}{3}} D.
\end{split}
\end{equation}
In the approximation $E_{\omega}/m_{\omega}\sim1$, we have
\begin{equation}
	g_d/g_s = \frac{1}{\sqrt{2}}\times\frac{1}{\frac{4}{3}Q^2 B_2(Q)}\times D/S,
\end{equation}
which gives $|g_d/g_s|=0.134\pm0.007$ and $\arg(g_d/g_s)=(10.5\pm4.6)^{\circ}$. In the following numerical calculations, we will take $g_s=1$ and $g_d=0.134 e^{i 10.54^{\circ}}$.

\subsection{Coherent cascade and spin sum}

Combining the helicity projections of $D\to b_1(1235) W^*$ in Eq.~\ref{eq:hadronic_helicity}, $W^*\to\ell^+\nu_\ell$ in Eq.~\ref{eq:weak_helicity}, and $b_1(1235)\to\omega\pi$ in Eq.~\ref{eq:strongdecay_helicity}, the decay amplitude of $D\to b_1(1235)\ell^+\nu_\ell, b_1(1235)\to\omega\pi$ is given as
\begin{equation}
\begin{split}
	\mathcal M_{\lambda_\omega,\lambda_\ell} &= \mathcal{N} h_{\lambda_\omega} \times \Bigl[c_- L_{\lambda_\ell,-1}\mathcal{D}_{-1,\lambda_{\omega}} \\
	& + c_+ L_{\lambda_\ell,+1}\mathcal{D}_{+1,\lambda_{\omega}} + \bigl(c_0 L_{\lambda_\ell,0}-c_t L_{\lambda_\ell,t}\bigr)\mathcal D_{0,\lambda_{\omega}}\Bigr].
\end{split}
\end{equation}
Here, the overall coefficient $\mathcal{N}=\frac{G_F}{\sqrt{2}} V_{cd}{\rm BW}_{A}(s_A)$ and the Wigner-$D$ function is defined as
\begin{equation}
	\mathcal{D}_{\lambda_{W},\lambda_{\omega}}\equiv \mathcal{D}^{1 *}_{\lambda_{W},\lambda_{\omega}}(\theta_A,\phi_A,0).
\end{equation}
The decay intensity is obtained by summing over the spins of $\ell^+$ and $\omega$, which is written
\begin{equation}
	\mathcal I\equiv \sum_{\lambda_{\ell}=\pm1/2}\sum_{\lambda_{\omega}=0,\pm1}|\mathcal M_{\lambda_\omega,\lambda_\ell}|^2.
\end{equation}
It is further decomposed according the polarization of $\omega$ as
\begin{equation}
\begin{split}
\mathcal I = |\mathcal{N} f_\ell|^2\times\bigl( |h_L|^2 P_L + |h_T|^2 P_T\bigr).
\end{split}
\end{equation}
Here, $|h_T|^2$ and $|h_L|^2$ are the intensity of the strong decay $b_1(1235)\to \omega\pi$ with transversely and longitudinally polarized $\omega$, respectively, which are given as
\begin{equation}
\begin{split}
|h_T|^2=&|g_s|^2+\frac{16}{9}Q^4B_2(Q)^2|g_d|^2+\frac83Q^2B_2(Q)\Ree(g_sg_d^*), \\
|h_L|^2=&\frac{E_\omega^2}{m_\omega^2}\Big[|g_s|^2+\frac{64}{9}Q^4B_2(Q)^2|g_d|^2\\
&-\frac{16}{3}Q^2B_2(Q)\Ree(g_sg_d^*)\Big]. \\
\end{split}
\end{equation}
The $P_L$ and $P_T$ are the angular dependence for the weak decays $D\to b_1(1235)\ell^+\nu_\ell, b_1(1235)\to\omega\pi$ with transversely and longitudinally polarized $\omega$, with their sum written as
\begin{equation}
	(P_L+P_T)=\Sigma\equiv\bigl(\Sigma_+ +\widehat m_\ell^2\Sigma_-),
\end{equation}
where
\begin{equation}
\begin{split}
\Sigma_+=&c_-^2(1+\cos\theta_\ell)^2+c_+^2(1-\cos\theta_\ell)^2+2c_0^2\sin^2\theta_\ell,\\
\Sigma_-=&(c_-^2+c_+^2)\sin^2\theta_\ell+2(c_0\cos\theta_\ell+c_t)^2.\\
\end{split}
\end{equation}
Without $b_1(1235)$ polarization accessed via $b_1\to\omega\pi$,  the decay intensity reduces to $\mathcal{I}\propto\bigl(\Sigma_+ +\widehat m_\ell^2\Sigma_-\bigr)$. Finally, the angular dependence of the longitudinal component \mbox{$P_L=F_0+F_1\cos\phi_A+F_2\cos2\phi_A$} is given as
\begin{equation}
\begin{split}
F_0=&\frac{\sin^2\theta_A}{2}
\Big[c_-^2(1+\cos\theta_\ell)^2
+c_+^2(1-\cos\theta_\ell)^2 \\
&+\widehat m_\ell^2\sin^2\theta_\ell(c_-^2+c_+^2)\Big]\\
&+2\cos^2\theta_A\Big[c_0^2\sin^2\theta_\ell+\widehat m_\ell^2(c_0\cos\theta_\ell+c_t)^2\Big], \\
F_1=&\sin2\theta_A\,\sin\theta_\ell\Big[c_0(c_--c_+)\\
&+(1-\widehat m_\ell^2)c_0\cos\theta_\ell(c_-+c_+)-\widehat m_\ell^2c_t(c_-+c_+)\Big],\\
F_2=&-(1-\widehat m_\ell^2)\sin^2\theta_\ell\sin^2\theta_A\,c_-c_+.
\end{split}
\end{equation}

\begin{figure*}[htbp]
	\centering
	\includegraphics[width=0.90\textwidth]{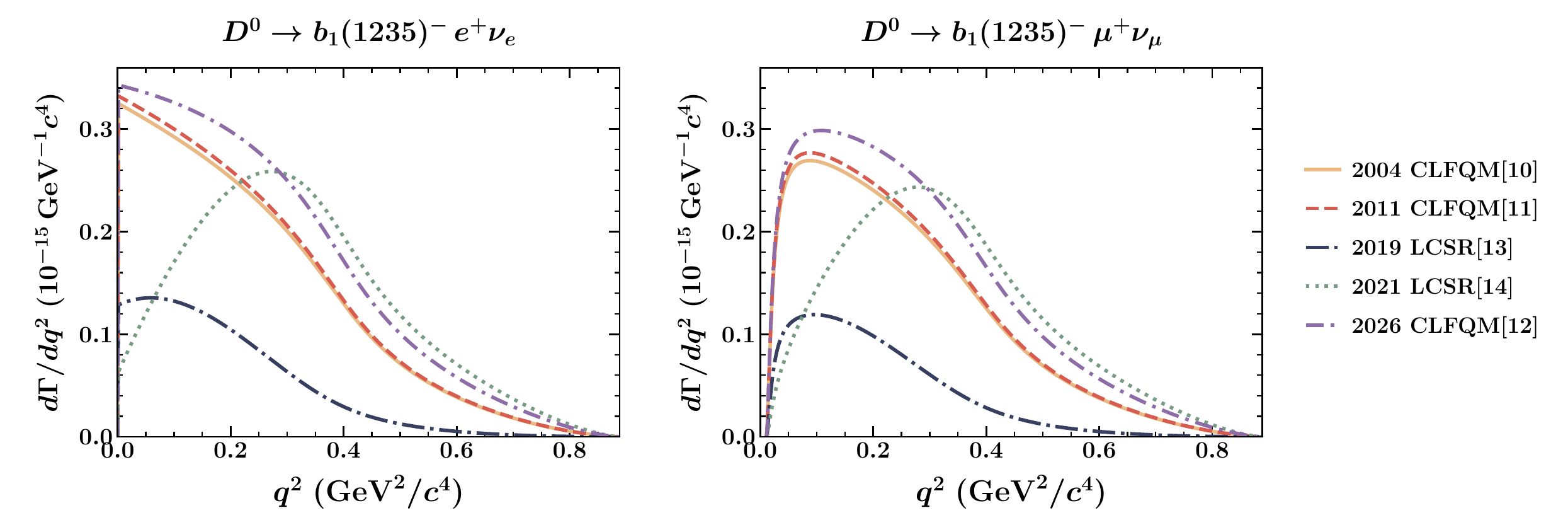}
	\caption{Predicted $q^2$ differential rates for
		$D^0\to b_1(1235)^-(\to\omega\pi^-)\ell^+\nu_\ell$,
		for $\ell=e$ (left) and $\mu$ (right), using five form factor predictions.}
	\label{fig:q2}
\end{figure*}

\subsection{Differential decay width}

The four-body phase space of $D\to \omega\pi \ell^+\nu_\ell$ can be decomposed into three two-body phase space factors as
\begin{equation}
\begin{split}
d\Phi_4(D\to \omega\pi \ell^+\nu_\ell)
&=\frac{ds_A\,dq^2}{(2\pi)^2}\times d\Phi_2(D\to b_1(1235)\ell^+\nu_\ell) \\
&\times d\Phi_2(b_1(1235)\to\omega\pi) \times d\Phi_2(W^*\to\ell^+\nu_\ell),
\end{split}
\label{eq:phasefactorization}
\end{equation}
where the convention of two-body phase factor $d\Phi_2(a\to bc) = \frac{Q}{16\pi^2M_a} d\Omega_a$ is used. Since the derived decay intensity is not dependent on $\phi_\ell$ and $\Omega_D$, these angles are integrated out. Hence, the total phase space factor is given as
\begin{equation}
	\frac{d^5\Phi_4}{ds_A\,dq^2\,d\cos\theta_A\,d\cos\theta_\ell\,d\phi_A} =\frac{X\beta_A\beta_\ell}{2(4\pi)^6m_D^2}.
\end{equation}
Here, $m_D$ is the mass of $D$ meson, and three kinematic factors are defined following the convention in Ref.~\cite{Lee:1992ih}:
\begin{equation}
\begin{split}
	X&=\sqrt{(k\cdot q)^2-s_Aq^2}
	=\frac12\sqrt{\lamfun(m_D^2,s_A,q^2)},\\
	\beta_A&={\sqrt{\lamfun(s_A,m_P^2,m_\omega^2)}}/{s_A},\\
	\beta_\ell&={\sqrt{\lamfun(q^2,m_\ell^2,0)}}/{q^2}
	=1-{m_\ell^2}/{q^2}.
\end{split}
\end{equation}
Finally, the differential decay width is given as
\begin{equation}
\label{eq:dd_width}
\frac{d^5\Gamma}{ds_A\,dq^2\,d\cos\theta_A\,d\cos\theta_\ell\,d\phi_A} =\frac{X\beta_A\beta_\ell}{4(4\pi)^6m_D^3}\mathcal{I}.
\end{equation}

\begin{figure*}[htbp]
	\centering
	\includegraphics[width=0.98\textwidth]{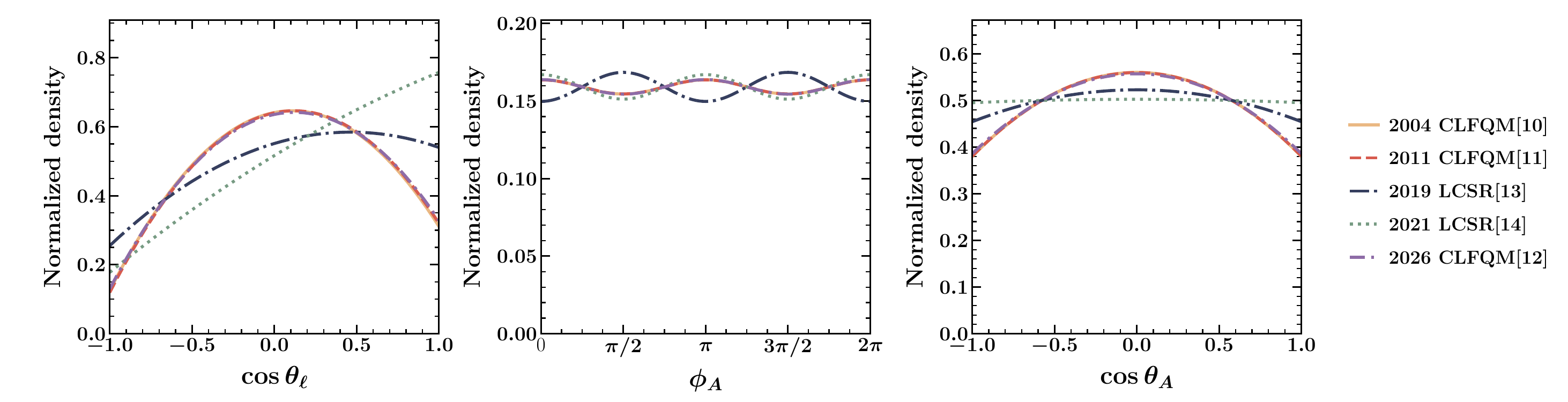}
	\caption{Normalized one-dimensional angular projections
		for $D^0\to b_1(1235)^-(\to\omega\pi^-)e^+\nu_e$:
		$\cos\theta_\ell$ (left), $\phi_A$ (middle), and
		$\cos\theta_A$ (right), using five form factor predictions.}
	\label{fig:angles}
\end{figure*}

\section{Branching fractions and lepton flavor universality ratios}
\label{sec:rates}

After integrating over all the kinematic variables in Eq.~\ref{eq:dd_width}, the total decay width $\Gamma_{\rm tot}$ and branching fraction $\mathcal{B}=\Gamma_{\rm tot}/\Gamma_{D}$ can be calculated numerically. The $q^2$-dependent form factors $A(q^2)$ and $V_{0,1,2}(q^2)$ of $D\to b_1(1235)\ell^+\nu_\ell$ predicted in the CLFQM~\cite{Cheng:2003sm,Verma:2011yw,Yang:2025gfz} and LCSR~\cite{Momeni:2019uag,Huang:2021owr} frameworks are used. Among these, no explicit $q^2$-dependence is given in Ref.~\cite{Huang:2021owr}, thus an effective single-pole form $F(q^2)=F(0)/(1-q^2/m_D^2)$ is assumed. Details about the predicted $q^2$-dependent form factors are given in Appendix~\ref{appendix}. No explicit form factor is given in ISGW papers~\cite{Isgur:1988gb,Scora:1995ty}, thus are not considered in the following numerical estimations.

In the following numerical estimations, only the results of $D^0\to b_1(1235)^-\ell^+\nu_\ell$ will be given, and their isospin partners $D^+\to b_1(1235)^0\ell^+\nu_\ell$ could be accessed by assuming isospin conservation, namely $d\Gamma_{D^0}=2\times d \Gamma_{D^+}$.

Table~\ref{tab:bf} summarizes the branching fractions of the positronic and muonic channels for $D^0\to b_1(1235)^-\ell^+\nu_\ell$, and their lepton flavor universality ratio $R_{\mu/e}=\Br_\mu/\Br_e$, based on various predicted form factors. Except for Ref.~\cite{Momeni:2019uag}, the other four calculated branching fractions show good agreement with the previous BESIII measurement result~\cite{BESIII:2024pwp}.

\begin{table}[htbp]
	\caption{Predicted branching fractions for $D^0\to b_1(1235)^-\ell^+\nu_\ell$ for positronic and muonic channels, and their lepton flavor universality ratio. The branching fractions are in units of $10^{-5}$. BESIII measurement is based on assumption $\Br(b_1(1235)\to\omega\pi)=1$}
	\label{tab:bf}
	\setlength{\tabcolsep}{8pt}
		\begin{tabular}{lccc}
			\hline
			\hline
			Form factors & $\Br_e$ & $\Br_\mu$ & $R_{\mu/e}$\\
			\hline
			2004 CLFQM~\cite{Cheng:2003sm} & 7.28 & 6.37 & 0.88\\
			2011 CLFQM~\cite{Verma:2011yw} & 7.47 & 6.55 & 0.88\\
			2019 LCSR~\cite{Momeni:2019uag} & 2.59 & 2.20 & 0.85\\
			2021 LCSR~\cite{Huang:2021owr} & 7.11 & 6.48 & 0.91\\
			2026 CLFQM~\cite{Yang:2025gfz} & 8.80 & 7.77 & 0.88\\
			\hline
			BESIII~\cite{BESIII:2024pwp} & $7.2\pm1.8^{+0.6}_{-0.8}$ & --- & --- \\
			\hline
			\hline
		\end{tabular}
\end{table}

\section{Partial decay width and angular distributions}
\label{sec:angles}

To discriminate between the four theoretical works with the similar predicted branching fractions, we further investigate their partial decay widths and angular distributions. By integrating over the other four kinematic variables shown in Eq.~\ref{eq:dd_width} numerically, partial decay widths $d\Gamma/dq^2$ and one-dimensional angular distributions $d\Gamma/dx$ with $x=(\cos\theta_\ell,\cos\theta_A,\phi_A)$ are determined based on various form factor predictions~\cite{Cheng:2003sm,Verma:2011yw,Yang:2025gfz,Momeni:2019uag,Huang:2021owr}.

As shown in Figs.~\ref{fig:q2} and~\ref{fig:angles}, the form factors calculated by Ref.~\cite{Huang:2021owr} result in distinct partial decay widths and angular distributions compared to the others. All three CLFQM works give consistent form factors at $q^2=0$ and slightly different $q^2$-dependence, thus only limited difference could be seen in their partial decay widths.

\section{Summary}
\label{sec:summary}

Using the helicity amplitude technique, the decay intensity for the charmed meson semileptonic decay
$D\to b_1(1235)\ell^+\nu_\ell, b_1(1235)\to\omega\pi$ with $\ell=e,\mu$, is derived. Combining the derived differential decay widths and theoretically predicted $q^2$-dependent form factors, we present numerical estimations on their branching fractions, lepton flavor universality ratios, partial decay widths, and angular distributions. Distinct branching fractions and differential decay widths are observed between various theoretical predictions, except three CLFQM works with similar form factors. Further amplitude analysis of $D\to b_1(1235)e^+\nu_e, b_1(1235)\to\omega\pi$ with the full $20.3$ fb$^{-1}$ data sample at $\sqrt{s}=3.773$~GeV at BESIII~\cite{BESIII:2024lbn} and forthcoming large data sample at super $\tau$-charm factory~\cite{Cheng:2022tog} is desired to discriminate these theoretical predictions, and potential extension to the muonic channels would provide a test on lepton flavor universality.

\bibliographystyle{apsrev4-1}
\bibliography{references}

\onecolumngrid
\appendix
\section*{Appendix}
\label{appendix}

Table~\ref{tab:form_factors} summaries the predicted $q^2$-dependent form factors of $D\to b_1(1235)$ in Refs.~\cite{Cheng:2003sm,Verma:2011yw,Yang:2025gfz,Momeni:2019uag,Huang:2021owr}.

\begin{table*}[htbp]
	\caption{\label{tab:form_factors}
		The predicted $q^2$-dependent form factors of $D\to b_1(1235)$. Each triplet gives $(F(0),a,b)$. With $x=q^2/m_D^2$, parametrization I is $F(q^2)=F(0)/(1-ax+bx^2)$ and parametrization II is $F(q^2)=F(0)/[(1-x)(1-ax+bx^2)]$. Dashes indicate that no corresponding parameters are given in Ref.~\cite{Huang:2021owr}.}
	\begin{ruledtabular}
		\begin{tabular}{lccccc}
			Form factor
			& $A$
			& $V_0$
			& $V_1$
			& $V_2$
			& Parametrization \\
			\hline
			2004 CLFQM~\cite{Cheng:2003sm}
			& $(0.11,1.08,0.54)$
			& $(0.49,0.89,0.28)$
			& $(1.37,0.46,0.05)$
			& $(-0.10,0.21,0.67)$
			& I; II for $V_2$ \\
			2011 CLFQM~\cite{Verma:2011yw}
			& $(0.12,1.09,0.50)$
			& $(0.50,0.98,0.26)$
			& $(1.39,0.44,0.05)$
			& $(-0.10,0.26,0.90)$
			& I; II for $V_2$ \\
			2019 LCSR~\cite{Momeni:2019uag}
			& $(-0.41,0.45,0.40)$
			& $(-0.32,-2.30,9.12)$
			& $(-0.22,1.20,1.47)$
			& $(0.21,-0.31,-0.38)$
			& I \\
			2021 LCSR~\cite{Huang:2021owr}
			& $(-0.24,\text{--},\text{--})$
			& $(-0.16,\text{--},\text{--})$
			& $(-1.78,\text{--},\text{--})$
			& $(-0.24,\text{--},\text{--})$
			& \text{--} \\
			2026 CLFQM~\cite{Yang:2025gfz}
			& $(0.12,0.77,0.21)$
			& $(0.50,0.58,0.12)$
			& $(1.40,0.43,0.11)$
			& $(-0.10,1.15,1.94)$
			& II \\
		\end{tabular}
	\end{ruledtabular}
\end{table*}

\end{document}